\documentclass[preprint,showpacs,secnumarabic,preprintnumbers,amsmath,amssymb,aps]{revtex4}

\usepackage{graphicx}
\usepackage{dcolumn}
\usepackage{bm}
\usepackage[dvipdfm,colorlinks=true,linkcolor=blue,citecolor=blue,bookmarks=false]{hyperref}%

\begin{document}

\preprint{APS/123-QED}

\title{Direct Measurement of Thermal Fluctuation of High-Q Pendulum}

\author{Kazuhiro Agatsuma$^{1,*}$}
\author{Takashi Uchiyama$^1$}%
\author{Kazuhiro Yamamoto$^2$}%
\author{Masatake Ohashi$^1$}%
\author{\\Seiji Kawamura$^3$}%
\author{Shinji Miyoki$^1$}%
\author{Osamu Miyakawa$^1$}%
\author{Souichi Telada$^4$}%
\author{Kazuaki Kuroda$^1$}%

\affiliation{%
$^1$Institute for Cosmic Ray Research, University of Tokyo, Kashiwa, Chiba 277-8582, Japan\\
$^2$Max Planck Institute for Gravitational Physics (Albert Einstein Institute),Callinstrasse 38, D-30167 Hannover, Germany\\
$^3$National Astronomical Observatory of Japan, Mitaka, Tokyo 181-8588, Japan\\
$^4$National Institute for Advanced Industrial Science and Technology, Tsukuba, Ibaraki 305-8563, Japan
}%

\date{\today}

\begin{abstract}
We achieved for the first time a direct measurement of the thermal fluctuation of a pendulum in an off-resonant region 
using a laser interferometric gravitational wave detector. 
These measurements have been well identified for over one decade by 
an agreement with a theoretical prediction, 
which was derived by a fluctuation-dissipation theorem. 
Thermal fluctuation is dominated by the contribution of resistances in coil-magnet actuator circuits. 
When we tuned these resistances, the noise spectrum also changed according to a theoretical prediction. 
The measured thermal noise level corresponds to a high quality factor on the order of $ 10^5 $ of the pendulum. 
\end{abstract}

\maketitle
\textit{Introduction.}\,---
Mechanical thermal fluctuation has been studied in fields that require ultra-precise measurements, 
because it gives the fundamental limits of the sensitivity, even in an off-resonant region. 
A rigid cavity for laser frequency stabilization is limited by the thermal noise~\cite{Numata2004}
; also, 
a laser interferometric gravitational wave (GW) detector~\cite{GWdetector} is one of the most representative type of apparatus. 
GW detectors consist of mirrors suspended by pendulums to bring them close to a state of free mass. 
The thermally excited mechanical vibrations of the mirrors themselves and suspensions are kinds of fundamental limits of GW detectors. 

Direct observations of the mirror thermal fluctuation have been performed 
by \mbox{Numata}~\cite{Numata2003} or \mbox{Black}~\cite{Black2004}.
They identified the thermal fluctuation using the fluctuation-dissipation theorem (FDT)~\cite{Callen1951,Callen1952}, 
which could evaluate the fluctuation from the dissipation in thermal equilibrium.
These experiments could not observe the thermal fluctuation of the pendulum 
because the seismic noise and the mirror thermal fluctuation had large amplitudes in their experiments. 
The seismic noise was larger than the pendulum thermal fluctuation below 100 Hz.
The mirror thermal fluctuation was also larger than the pendulum thermal fluctuation above 10 Hz, 
because they set up a small spot size 
(radii of 49 $\mu\rm{m}$ and 85 $\mu\rm{m}$ of \mbox{Numata}~\cite{Numata2003} or 160 $\mu\rm{m}$ of \mbox{Black}~\cite{Black2004} on mirrors) 
so as to increase the mirror thermal fluctuation. 
Nobody has ever identified the thermal fluctuation of a pendulum in an off-resonant regime. 

Our experiment could achieve measurements of the pendulum thermal fluctuation 
in both off-resonant and in a wide-band region of over one decade, 
which is close to 
the most expected band of GW detection (around 100 Hz).
The seismic noise could be reduced by a quiet environment at 1000 m underground site.
The mirror thermal noise was also reduced owing to a large spot size 
(radii of 4.9 mm and 8.5 mm, where the electric field fell off to 1/e) of a laser on a mirror surface. 
The quality factors of our pendulums using 1.8 kg mirrors were about on order of $ 10^{5} $, 
which were no less high than that of the current km-scale interferometer pendulums~\cite{LIGO1999,VIRGO1997}. 
As a manner of identification of the thermal fluctuation of the pendulum, 
we made use of coil-magnet actuators, which had some merits. 
One was the dissipation that were occurred in coil-circuits, was easy to control and to analyze. 
Another was that all of the parameters were measurable. 
The other was that they did not cause any modification of the interferometer components with changes of the dissipation in the coil-circuits.
These merits allowed a high reliability to our experiment. 

This Letter describes direct measurements of the thermal fluctuation of the pendulum, 
and compares it with predicted estimates from the FDT over a wide band from 20 Hz to 400 Hz.
We observed shifts in the noise floors of the interferometer 
when we controlled the dissipation of a pendulum by exchanging the resistances of the coil-magnet actuators. 
The measured spectra and their shift agreed with a theoretical prediction. 
This experiment was the first to identify the thermal fluctuation of a pendulum in the off-resonant frequency regime.

\textit{Theoretical background.}\,---
Pendulum thermal fluctuation is caused by several kinds of dissipations in the whole pendulum. 
There have been experiments and studies so far for each type of dissipation: 
an internal loss in the materials of suspension wires~\cite{Saulson1990,Gonzalez1995},
clamps of wires~\cite{Dawid1997}, 
residual air~\cite{Kajima1999},
and
the bobbins or holders of coils~\cite{Cagnoli1998,Frasca1999}. 
Thermal fluctuation caused by coil-magnet actuators differs from other dissipations
from the viewpoint of easy tuning and the well-known frequency dependence.

In the current interferometric GW detectors that have a Fabry-Perot (FP) cavity in their Michelson arms, 
coil-magnet actuators are equipped to keep the FP cavity on resonance by controlling the mirror positions. 
One actuator consists of a pair of a magnet glued onto a mirror, and a coil-circuit 
connected to a coil-driver circuit. 
The dissipation in the coil-circuits of the actuators produces the pendulum thermal fluctuation  
because a pendulum is unified with coil-circuits, 
and is in thermal equilibrium with them due to coil-magnet coupling. 
An oscillation of the pendulum causes eddy currents by electromagnetic induction in the coil-circuits. 
These currents generate Joule heat in the resistances of the circuits. 
The pendulum thermal fluctuation can be calculated by applying the FDT to the dissipation from this Joule heat. 
The other evaluation method is to calculate the pendulum fluctuation 
caused by the thermal current of the resistance~\cite{Nyquist1928} via coil-magnet coupling.
These two estimates are consistent as long as the thermal equilibrium between the coil-circuits and the pendulum is maintained. 

As a model of the pendulum, we assume a suspended mirror (test mass of $ m $) 
and actuators constructed with a pair of coil-magnets of $ N $. 
When a current $ I $ flows inside of the coil-circuits, a force of $ F = I \alpha N $ is applied to the mirror,
where $ \alpha $ is the coupling factor per one coil-magnet actuator. 
That motion equation of the mirror in a frequency domain is written as
\begin{equation}
  m(-\omega ^2 + \omega _0^{2}) \tilde {x} + i \frac{N \alpha ^{2} \omega }{Z}  \tilde {x} = 0 . \label{eq:motion}
\end{equation}
Here, $ \omega _0 = 2 \pi f_0$, where $ f_0 $ is the resonant frequency of the pendulum, 
$ \tilde {x} $ is the displacement of the mirror and $ Z $ is the impedance of the coil-circuit, 
which is the sum of the coil's impedance itself, and the connected driver output impedance. 
It is assumed that the magnitude of the impedance in the coil-circuit is dominated by its resistance, $ R $; 
it can then be regarded as $ Z = R $. 
From the imaginary part of Eq.(\ref{eq:motion}), the quality factor is defined as
\begin{equation}
  Q = \frac{m \omega _0 R}{N \alpha ^2} .    \label{eq:Qc}
\end{equation}
The fluctuation is characterized by the loss angle, 
$ \phi = \omega /( \omega _0 Q ) $, 
which indicates viscous damping caused by an eddy current in the coil-circuits.
By using the FDT, the fluctuation of a harmonic oscillator i.e. a pendulum 
in a higher off-resonant region, is approximately expressed as 
\begin{equation}
  G = \frac{4 k_B T N \alpha ^2}{m^2 \omega ^4 R} .  \label{eq:Gc}
\end{equation}
$ \sqrt{G} $ is a one-sided power spectrum density,
$ k_B $ is the Boltzmann constant,
and $ T $ is the temperature of the suspension and the coil-circuits. 

\begin{figure}
\includegraphics[width=8.6cm,clip]{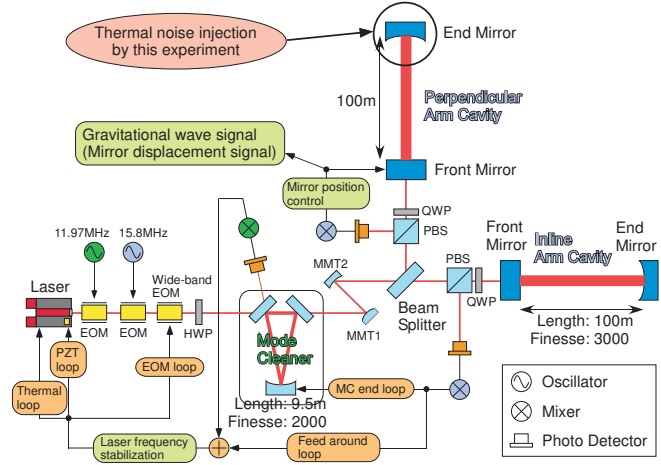}
\caption{\label{fig:CLIO} (color online). Schematic view of CLIO and its control system; 
CLIO is a so-called locked Fabry-Perot interferometer, 
which has 100 m Fabry-Perot cavities as Michelson arms and a mode cleaner. 
The optics, like the lenses, the Faraday-isolators, and some wave plates are omitted in this figure.
Abbreviated words mean: 
EOM, electro-optic modulator; 
HWP, half wave plate; 
QWP, quarter wave plate; 
MMT, mode matching telescope; 
and PBS, polarized beam splitter.
}
\vspace{-9pt}
\end{figure}
\textit{Experimental setup.}\,---
A GW detector, called CLIO (Cryogenic Laser Interferometer Observatory)~\cite{Miyoki2004}, 
was employed as a mirror displacement sensor. 
CLIO is a prototype for the next Japanese GW telescope project, 
LCGT (Large scale Cryogenic Gravitational-wave Telescope)~\cite{LCGT2006}, 
featuring the use of cryogenic mirrors and a quiet underground site. 
The main purpose of CLIO is to demonstrate a reduction of the mirror thermal noise by cooling sapphire mirrors. 
In this Letter, however, CLIO is described as being operated at room temperature. 
CLIO is located in the Kamioka mine, which is 220 km away from Tokyo, Japan, and lies 1000 m underground from the top of a mountain. 
This site is suitable for precise measurements of the displacement at below the 100 Hz region, 
because the seismic noise is smaller than that in an urban area by about 2 orders~\cite{Araya1993}. 
Through the progress of noise hunting in CLIO, 
the background (BG) noise has reached about $ 1 \times 10^{-18} $\,m/$ \sqrt{\rm{Hz}} $ at 100 Hz.
This sensitivity is close to the design sensitivity of $ 5 \times 10^{-19} $\,m/$ \sqrt{\rm{Hz}} $. 

Figure~\ref{fig:CLIO} shows a schematic view of CLIO. 
A laser beam, which has a power of 2 W and a 1064 nm wavelength (Innolight Inc. Mephisto), is used as a light source. 
It is shaped into TEM00 through a mode cleaner (MC) cavity with a length of 9.5 m, 
and then enters two 100 m length FP cavities after being divided by a beam splitter. 
These cavities are arranged in an L-shape. 
The cavities are kept on resonance (we call this state locked) by servo systems, 
which readout displacement signals of the mirrors using the Pound-Drever-Hall method~\cite{Drever1983}. 
The modulation frequencies are 15.8 MHz for the arm cavities and 11.97 MHz for the mode cleaner.
A multistage control system~\cite{Nagano2003} is applied for a laser frequency stabilization, 
which has two cascaded loops of the MC and an inline arm cavity.
The inline arm is locked by controlling the frequency of the laser
and the perpendicular arm is locked by controlling the position of the front mirror by using coil-magnet actuators.
From a feedback signal to the coil-magnet actuators, 
a differential displacement between two arm cavities, 
which corresponds to a GW signal, can be obtained. 

The four mirrors of the two arm cavities are individually 
suspended by 6-stage pendulums, 
which include 4-stage blade springs and 2-stage wire suspensions 
for isolation from any seismic vibration.
The mirrors are made of sapphire, 
whose weight is 1.8 kg, suspended by bolfur wires at the last stage. 
The last stage pendulum has a resonant frequency of about 0.8 Hz. 
Coil-magnet actuators are set for two mirrors in a perpendicular cavity. 
In order to observe the thermal fluctuation of the pendulum, 
we injected sufficient thermal noise to the end mirror (see Fig.~\ref{fig:CLIO}), 
so that its fluctuation could dominate, 
by controlling the dissipation at the coil-circuits. 
On the other hand, the actuator of the front mirror is used to keep the cavity locked. 

\begin{figure}
\includegraphics[width=6.3cm,clip]{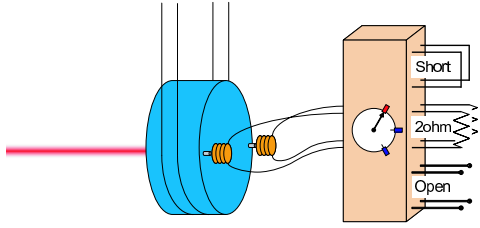}
\caption{\label{fig:setup} (color online). Schematic view of our experiment. 
The pendulum is composed of a sapphire mirror suspended by bolfur wires. 
Two coil-magnet actuators are set. 
The fluctuation of the pendulum is measured using a laser interferometer (CLIO). 
The resistances of the coil-circuits are changed by switching a relay-circuit to ``Short", ``2ohm", and ``Open". 
Therefore, we can control the injected thermal noise using this relay-circuit.
}
\vspace{-9pt}
\end{figure}

The coil-magnet actuator at the end mirror consists of two Nd-Fe-B magnets, 
which are glued onto the mirror 
and of two copper coils approached to magnets in the horizontal direction (see Fig.~\ref{fig:setup}).
The Nd-Fe-B magnets have a cylindrical shape.
This diameter is 2 mm and the length is 10 mm.
A copper wire of 0.5 mm diameter is used for solenoidal coils, 
and is wound by 15 turns times 6 layers around a ceramic bobbins, 
whose diameter is $ 16.5 $ mm.
In order to verify our measurement of the thermal fluctuation, 
the resistance $ R $ of Eq.(\ref{eq:Gc}) is changed as a variable parameter. 
$ R $ is the sum of the coil's impedance and a connected circuit impedance. 
Although a coil-driver circuit is usually used as the connected circuit, 
a relay-circuit including resistances is used to change $ R $ in this case. 
The relay-circuit is switched to three kinds of resistances, 
which are called ``Short", ``2ohm", and ``Open". 
The sum of the resistances of the coil-circuits and a relay-circuit were measured as 0.72 $ \Omega $ in ``Short", 
and 2.77 $ \Omega $ in ``2ohm". 
``Open" means that the end of the relay-circuit is opened, and then $ R $ becomes infinity. 
A small $ R $ causes a large thermal fluctuation for the pendulum from Eq.(\ref{eq:Gc}). 

It is necessary to know the quantities in Eq.(\ref{eq:Gc}) to estimate the thermal fluctuation. 
In our experiment, $ N $ is 2, $ m $ 1.8 kg, $ f_0 $ 0.8 Hz and $ T $ $ 3.0 \times 10^2 $ K.
The coupling factor, $ \alpha $, is an actuator force applied for the mirror per driver output current. 
That is yielded from measured actuator efficiency,
a resistance of volt-current conversion in the coil-driver, and the weight of a test mass. 
The actuator efficiency is a transfer function from the driver input voltage to the mirror displacement. 
The coil-magnet actuator efficiency at the front mirror is measured using a Michelson interferometer constructed with front mirrors and the BS.
The driver output current is obtained by a resistance of volt-current conversion, 
$ R_{\rm{c}} $ ($ R_{\rm{c}} = 50$ $ \Omega $ in CLIO). 
Using a measured actuator efficiency at 100 Hz as $ A_{100} $~\cite{alpha100}, 
the coupling factor per one coil-magnet actuator is 
$ \alpha = A_{100} R_{\rm{c}} m (2 \pi \times 100)^2 / N $. 
The efficiency of the end mirror actuator is measured by a calibrated actuator of the front mirror. 
Only during this calibration, the relay-circuit is exchanged to a driver-circuit. 
The value of $ \alpha $ at the end mirror is $ 6.9 \times 10^{-3} $ N/A. 

In this experiment, 
other dissipations of pendulums are adequately suppressed below the additional dissipation of the coil-circuits, 
so that the thermal fluctuation from the coil-magnet actuators at the end mirror can be observed at around 100 Hz. 
Coil bobbins are made of a macor (ceramic) which has an electrical conductivity of $ 10^{-13}$ $\rm{\Omega m} $. 
The pendulum is housed in a vacuum chamber, whose vacuum level is $ 10^{-3} $ Pa, so as to reduce any dissipation from residual air. 
The suspension thermal noise from a wire material was calculated to be below the level of the BG noise. 
The thermal noise from the coil-magnet actuator at the front mirror is under the BG noise, 
because the output impedance of the coil-driver is $ 10^4 $ $ \Omega $ at 100 Hz 
and $ \alpha $ is about ten-times smaller than that of the end mirror. 

\begin{figure}[t]
\includegraphics[width=8.6cm,clip]{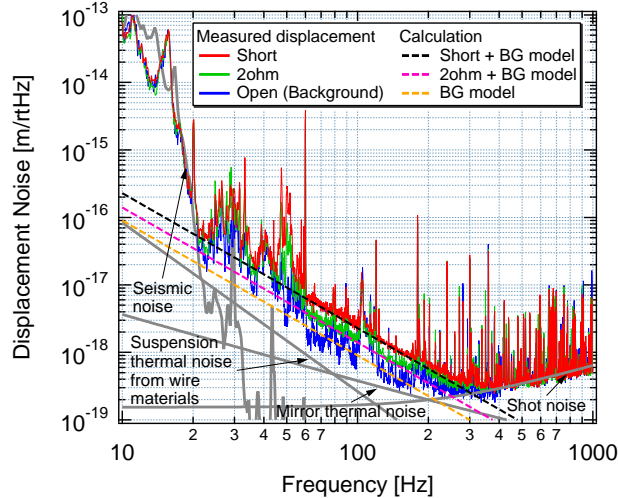}
\caption{\label{fig:all} (color). Measured spectra of the pendulum thermal fluctuation and their theoretically expected lines. 
The solid lines of red, green and blue denote the conditions of the ``Short", ``2ohm", and ``Open", respectively. 
The spectrum at ``Open" corresponds to the background (BG) noise of the interferometer. 
The doted lines indicate the sum of the theoretical calculation of the thermal fluctuation and the BG model.
The sensitivity is limited by the shot noise above 400 Hz, 
and by the seismic noise below 20 Hz.
The mirror thermal noise and the suspension thermal noise (from material dissipation) were calculated to be under the BG noise.
}
\vspace{-3pt}
\end{figure}
\textit{Results and discussion.}\,---
Figure~\ref{fig:all} shows measured spectra of the mirror displacements when we changed the resistances of coil-circuits, 
and also shows other noises of the detector. 
The noise levels from 20 Hz to 400 Hz are shifted by switching the dissipation. 
The measured values agree with the sums of theoretical estimates from Eq.(\ref{eq:Gc}) and the BG model. 
When the ``Open" was chosen, 
a BG noise of about $ 1 \times 10^{-18} $\,m/$ \sqrt{\rm{Hz}} $ at 100 Hz was measured, 
since the injection thermal noise was lower than the BG level, owing to a large amount of resistance, $ R $. 
For simplicity of calculations, the BG noise was modeled as $ 0.9 \times 10^{-18} $\,m/$ \sqrt{\rm{Hz}} $ at 100 Hz 
with a slope of $ f^{-2} $. 
In the case of ``Short", for instance, 
the noise level at 100 Hz was predicted to have a value of $ 2.3 \times 10^{-18} $\,m/$ \sqrt{\rm{Hz}} $, 
which includes a pendulum thermal fluctuation of $ 2.1 \times 10^{-18} $\,m/$ \sqrt{\rm{Hz}} $ and a value of the BG model. 
In ``2ohm", the total value of $ 1.4 \times 10^{-18} $\,m/$ \sqrt{\rm{Hz}} $ at 100 Hz included a thermal fluctuation of 
$ 1.1 \times 10^{-18} $\,m/$ \sqrt{\rm{Hz}} $ and the BG model. 
We also estimated Q of the pendulum, which corresponded to these thermal noises from Eq.(\ref{eq:Qc}). 
The calculated Q is $ 6.7 \times 10^4$ in ``Short" and $ 2.6 \times 10^5 $ in ``2ohm". 
A Q of $ 10^5 $ is as large as that of the pendulums in km-scale interferometers~\cite{LIGO1999,VIRGO1997}.

\begin{figure}
\includegraphics[width=7cm,clip]{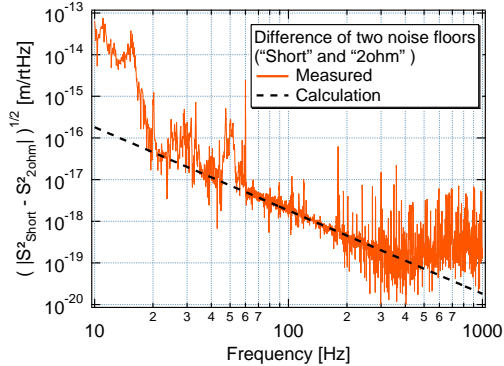}
\caption{\label{fig:difference} (color online). 
Comparison of the pendulum thermal fluctuation with a theoretical calculation by removing the BG noise; 
The orange line shows the thermal fluctuation difference of the ``Short" from the ``2ohm". 
The dash line shows the theoretical calculation corresponding to this difference. 
}
\vspace{-2pt}
\end{figure}
For a more precise and wide-band comparison between the measured spectra and theoretical calculations,
we show Fig.~\ref{fig:difference}, 
which is the difference between ``Short" and ``2ohm" 
taken as $ S_{\rm{d}} = ( \mid S_{\rm{Short}}^{\,2} - S_{\rm{2ohm}}^{\,2} \mid ) ^{1/2} $; 
here, $ S_{\rm{Short}},\, S_{\rm{2ohm}} $ indicate the spectra of ``Short", ``2ohm" in Fig.~{\ref{fig:all}}, respectively.
$ S_{\rm{d}} $, where the BG noise of two noise floors are canceled, 
is useful for comparing with the theoretical calculation of the thermal fluctuation. 
The result indicates a good agreement between the measurement and a calculation from 20 Hz to 400 Hz. 
Especially, the agreement is better at a region from 60 Hz to 300 Hz.
The peak structure around 50 Hz comes from the resonant motion of a coil holder at the end mirror excited by the seismic motion.
The noisy structure around 30Hz comes from the BG noise. 

\textit{Summary.}\,---
We identified for the first time the thermal fluctuation of a pendulum 
in an off-resonant regime and a wide-band range around the GW-band. 
This result was confirmed by a comparison 
between the measured thermal noises and a calculation from the FDT 
under tuning the resistance of coil-magnet actuators, 
which are the main sources of thermal noise.
The noise level indicates that the pendulums of CLIO have a high quality factor on the order of $ 10^{5} $. 
This experimental result is significant for fundamental physics 
and fields of a super-precise measurement, like GW detectors. 

This work was supported in part by
GCOE for Phys. Sci. Frontier, MEXT, Japan
and in part by a JSPS Grant-in-Aid for Scientific Research (No. 18204021).
\vspace{-6pt}



\begin{thebibliography}{99}
\bibitem{alpha100} 
The measurement at 100Hz is a middle point of our interest 
and an accurate measurement of the transfer function is possible.
\end{thebibliography}
\end{document}